# On the Dynamics of Imploding and Exploding Spherical Shock Wave Inside a Shock Tube

Saranyamol V. S.[1]     Talluri Vamsi Krishna.[2]     Mohammed Ibrahim S.[3]

Hypersonic Experimental aerodynamics Laboratory (HEAL)
Department of Aerospace Engineering,
Indian Institute of Technology, Kanpur, Uttar Pradesh, India, 208016.

**Abstract**

The present work aims to study the phenomenon of shock wave focusing and the effect of viscosity in it. The focusing is achieved with a shock tube and a converging section attached to it. The converging section transforms the planar shock into a spherical shock and focuses it into a confined area. A shock of an initial strength $M_s = 2.94$ has been chosen for the present studies. A detailed numerical study of the focusing region shows the formation of a mushroom-shaped structure behind the reflected shock and vortex formation. This was visualised through numerical shadowgraph images and by tracing the streamlines in the flow field. A study on the variation in temperature is carried out in order to have a quantitative assessment. It was found that the temperature inside the mushroom structure is higher than that behind the reflected shock. The study of species mass fraction in this region is also made. The flow inside the mushroom structure was found to be a reactive mixture of gas slug.

## 1 Introduction

Shock waves are strong and thin discontinuous regions that cause an abrupt rise in fluid properties like pressure, temperature, etc. Converging these high-energy shock waves to a tiny region of space will result in a very high energy concentration. This phenomenon is called shock wave focusing, and it has various applications like inertial confinement fusion [1][2], shock wave lithotripsy, material science [3], shock focusing ignition techniques [4], etc. As the strength of the shock increases, the focused region will encounter a high enough temperature that the gas in this region will start radiating [5]. Shock focusing resembles the phenomenon of gas bubble sonoluminescence [6], supernova collapse [7] [8], etc., and also finds its place to explore the study in Richtmyer–Meshkov instability [9].

---

[1]PhD Scholar: saranya@iitk.ac.in
[2]PhD Scholar: tkrishna@iitk.ac.in
[3]Corresponding Author: Assistant Professor, ibrahim@iitk.ac.in

The shock focusing phenomenon has been of great interest to researchers since 1942. Guderly [10] was the first to do theoretical studies on strong cylindrical and spherical shock waves. He proposed a self-similar solution in an ideal gas flow, which express the radius of converging shock wave as a function of time. Pioneer experimental study on converging shock waves was done by Perry and Kantrowitz [11], achieving cylindrical shock focusing with a tear-drop insert inside a shock tube. Several other shock focusing techniques like hemispherical implosion chamber [12], parabolic reflector in shock tube [13], annular shock tube [14], etc., were achieved later.

The challenge of producing a spherical shock focusing with the help of a shock tube was successfully overcome by Apazidis et al. [15]–[18]. A perfectly contoured converging section helped to smoothly vary the shock profile from planar to spherical contour with minimum diffusion losses to the shock. They reported that experiments with argon as test gas resulted in a temperature of 27,000 K in the focusing region [15].

Spherical shock wave focusing in 'air' test gas is not studied in much detail. Keeping this in mind, an experimental campaign followed by numerical investigations is carried out to understand the flow dynamics in the shock-focused region. Experiments are carried out to focus a planar shock of initial strength $M_s =2.94$. Unsteady pressure measured throughout the experiments is compared with numerical simulations for validation. Further understanding of the phenomenon is made through numerical visualizations and a detailed study of thermodynamic properties and species distribution during and after shock focusing. The influence of flow viscosity on the focusing phenomenon is also studied by comparing inviscid and viscous flow simulations.

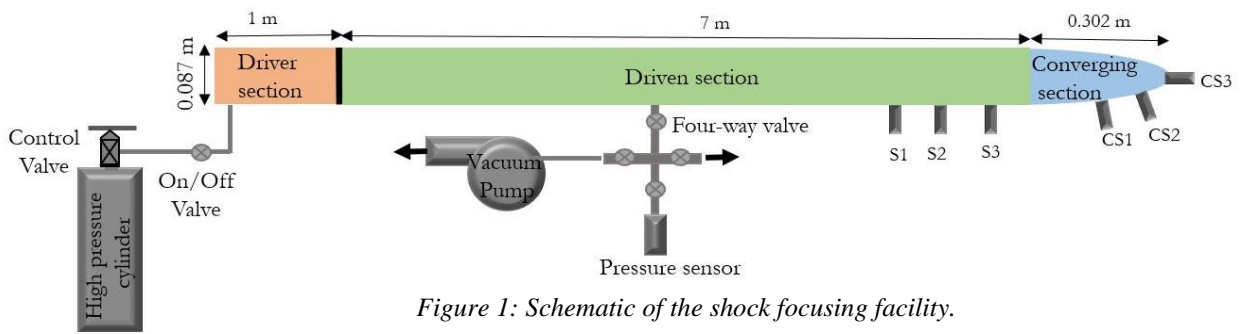

*Figure 1: Schematic of the shock focusing facility.*

## 2  Experimental Methodology

### 2.1  Shock Tube Facility

The current experiments are performed using the shock tube facility, 'S1 (Vaigai)' at the Hypersonic Experimental Aerodynamics Laboratory (HEAL), Indian Institute of Technology, Kanpur, India. The facility has a 1 m long driver section, a 7 m long driven section, and an 87 mm internal diameter. The schematic of the facility is shown in Figure 1. As shown in the figure, a converging section of length 302 mm is attached at the driven section end. For the current test, the initial fill condition in the driven section was 0.025 MPa generating a shock of speed $M_s = 2.94$. The test gas used in driven section was air, and the gas filled in the driver section was helium. An aluminium diaphragm separated the driver and driven sections. The temperature throughout the shock tube was unaltered, and the room temperature during the experiments was 300 K.

### 2.2  Shock Focusing facility

The converging section attached to the shock tube transforms the planar shock generated in the shock tube into a smooth spherical shock with minimum loss to the shock. The internal diameter of the tube reduced gradually from 87 mm to 18 mm at the focusing end wall. The design of the converging section was made according to the geometric relations proposed by Malte [19]. Unsteady pressure measurements are carried out using ICP® pressure sensor of PCB piezotronics, model No-113B22, flush-mounted along the surface of the facility, as can be seen in Figure 1. The pressure data was acquired using NI-USB6356, a multifunction I/O DAQ device at a rate of 1.0 Mega samples per channel over a duration of 0.25 seconds. The sensors were connected to the DAQ through a PCB signal conditioner (Model No. 482C05). The uncertainty of the unsteady pressure is ±10% about the measured values [20][21]

*2.3   Numerical methodology*

Numerical simulations are carried out using commercially available Computational fluid dynamics (CFD) software ANSYS-Fluent® version 2021-R1. A density-based transient implicit solver is used with an AUSM flux type scheme. In spatial discretization, the second-order scheme is used to resolve the flow, and Green-Gauss node-based model is used to determine gradients. The second-order implicit transient formulation is used to get better accuracy in the transient behaviour of moving waves.

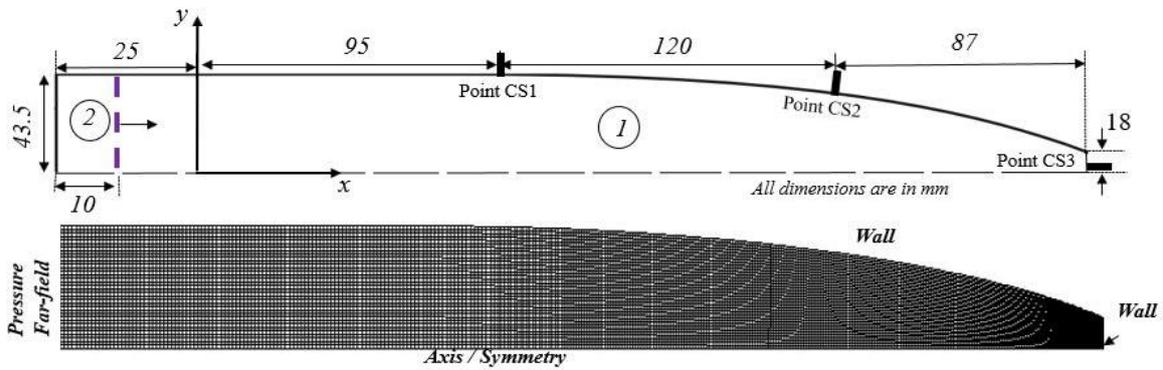

*Figure 2: Domain and boundary conditions for computational simulations. The location of pressure monitored are also shown in figure as CS1, CS2 and CS3.*

The axisymmetric domain used for the current simulation is shown in Figure 2. The converging section is 302 mm long. An additional 25mm constant-area section is added ahead of the domain's converging section, which resembles the flow inside the shock tube. The initial 10 mm of this constant-area region is patched with the conditions behind the incident shock. The initial fill conditions of driven gas are marked as region (1), and the conditions behind the incident shock are marked as region (2) in the figure. The pressure is monitored throughout the simulations at three locations. The distance between these three points is the same as that on the shock-focusing facility.

The distribution of the mesh and the boundary condition used is also shown in Figure 2. The Pressure far-field boundary condition is given to the inlet. Axis boundary condition helps produce the axisymmetric domain, and the other boundaries provided are the wall. All wall boundaries use slip conditions for inviscid flow, where there is no friction between fluid and surface and no-slip conditions for viscous flow.

Structured quadrilateral spatial grids are generated for 2D axisymmetric simulations. Grid independence studies have been carried out, and the result is shown in Figure 3. The pressure monitored at the focusing end wall for all the mesh configurations is compared with the experimental pressure value for the same test conditions ($M_s = 2.94$). Grid with 19160 mesh faces is found to be sufficient for the present analysis and is finalized. The analysis is done at a time step of $1 \times 10^{-7}$ with 60 iterations per time step [22]. At every time step in the scaled residuals, an absolute convergence criterion of $10^{-9}$ is achieved.

*Table 1: Initial mass fraction distribution of species*

| Species | Mass fraction |
|---|---|
| $N_2$ | 0.76 |
| $O_2$ | 0.23 |
| ar | 0.01 |
| NO | 0 |
| N | 0 |
| O | 0 |

The viscous model used is kω-SST since it is suitable for flow with chemical reactions and near-wall dissociation [23]. High-temperature effects like temperature-dependent $C_P$ variation and temperature-dependent species distribution are included in all the simulations [24] [22]. Since the test gas used for the current study is air, six species ($N_2$, $O_2$, NO, N, O) and the associated 11 reaction model (including third body reactions) [25] are added to the simulations. The initial mass fraction value of each species is mentioned in table 1. The finite rate model was considered as the turbulence chemistry interaction model for computing the reaction rates while solving the species transport equation. Details of reactions used in the current simulation are listed in table 2.

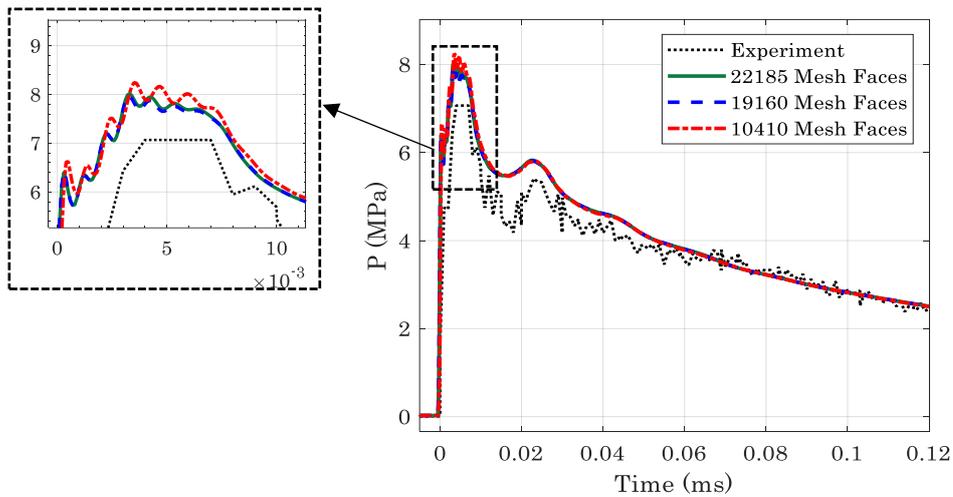

*Figure 3: Results of Grid Independence study showing the comparison of pressure monitored at a point on the focusing end wall.*

*Table 2: Chemical reactions included for simulations*

| No. | Reaction name | Forward Reaction |
|---|---|---|
| 1 | $O_2$ dissociation | $O_2$ + M → 2O + M(N) |
| 2 | $N_2$ dissociation | $N_2$ + M → 2N + M(O) |
| 3 | NO dissociation | NO + M → N + O + M ($O_2$) |
| 4 | $N_2$-O Exchange | $N_2$ + O → NO + N |
| 5 | NO- O Exchange | NO + O → N + $O_2$ |
| 6 | $N_2$-N Exchange | $N_2$ + N → 3N |
| 7 | $O_2$-O Exchange | $O_2$ + O → 2O + O |
| 8 | $O_2$- $O_2$ Exchange | $O_2$ + $O_2$ → 2O + $O_2$ |
| 9 | $O_2$-$N_2$ Exchange | $O_2$ + $N_2$ → 2O + $N_2$ |
| 10 | $N_2$- $N_2$ Exchange | $N_2$ + $N_2$ → 2N + $N_2$ |
| 11 | NO dissociation | NO + M → N + O + M(O) |

The forward reaction rate for these reactions is effectively estimated using the Arrhenius equation (equation 1), where the constant parameters are obtained from the literature [25].

$$K_f = AT^\beta e^{-E/RT} \qquad \text{equation 1}$$

Where, $K_f$ : Forward rate constant, A: Pre-exponential factor, β: Temperature exponent, E: Activation energy for the reaction, and R: Universal gas constant.

## 3  Results and Discussion

The overall observations made for focusing a shock of initial strength $M_s$=2.94 is described in this section. The section is divided into four subsections. The first subsection, 3.1, compares the pressure histories when the shock traverses through the converging section obtained experimentally and numerically. Once the reliability of the numerical simulations is established, a further in-depth study is carried out with the help of numerical simulations. The second subsection, 3.2, includes the observations made on the shock focusing phenomenon with the help of numerical shadowgraph images. A mushroom-shaped structure is established behind the shock after reflection from the focusing end wall. A detailed study on the distribution of temperature and species mass fraction inside the mushroom structure is also discussed in this subsection. A comparison of inviscid and viscous flow simulations is also presented throughout the subsections, which enables a study of the viscosity effect.

*3.1    Comparison of Pressure data*

The comparison of pressure distribution obtained through experiments and simulation at the locations CS1, CS2, and CS3 are shown in Figure 4. The time at which the shock reaches sensor CS3 is taken as zero for reference. The scale of the x-axis is adjusted in each image for a better display of the corresponding signal. The pressure signal in CS1 shows a sharp rise when the incident shock arrives, and the pressure remains in a constant value until the reflected shock reaches that location. The second rise corresponds to the arrival of the reflected shock after focusing.

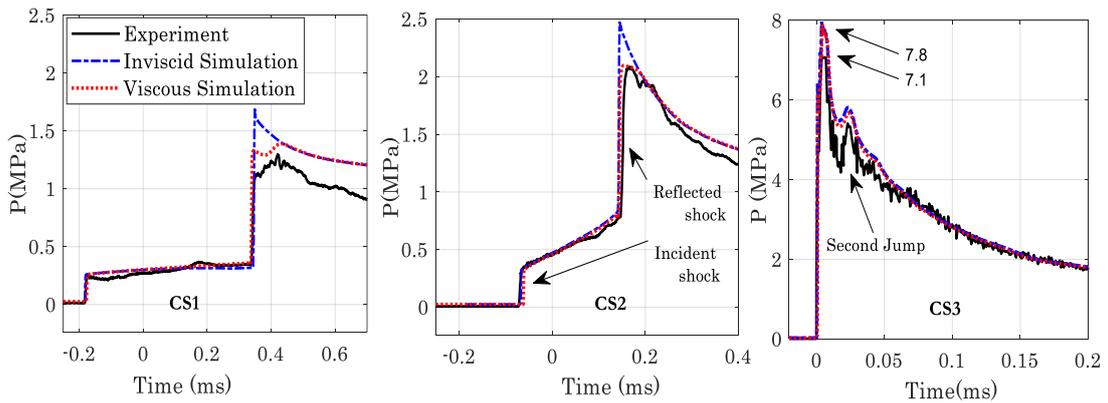

*Figure 4: Pressure measurements in all three sensor locations, CS1, CS2, and CS3, inside the converging section. Incident and reflected shocks can be seen in CS1 and CS2. Whereas CS3 shows a sudden rise in the pressure value*

At location CS2, the pressure value increases gradually behind the incident shock and drops gradually behind the reflected shock due to the area change. Comparing pressure signals at CS1 and CS2, the viscous simulation shows better agreement with experimental measurements. Since the monitor points are on the wall, the boundary layer influences the pressure signal, where the dominant viscous effects are present.

The incident shock hits CS3 and reflects from this point. The signal shows a sudden and sharp pressure rise and rapidly falls to an equilibrium value in a very short duration of time (0.2 ms). Both experiments and simulations follow the same trend of the falling phase. However, there is a 9.4 % difference between the peak pressure values of experiments and simulations. A second jump is observed in all the pressure signal of CS3 at time 24 µs after focusing. The possible explanation for the second pressure jump is given in the next subsection.

*3.2    Numerical Characterisation*

While passing through the converging section, the planar shock transforms into a spherical shock. The shape of the shock can be clearly monitored with the help of density contour images, as presented in our previous work [22]. Other visualization techniques like shadowgraph method show the derivatives of density, which helps to better analyse the shock and the flow field behind it. The instantaneous shadowgraph image showing the shape of the shock and the flow physics happening in the focusing region behind the reflected shock for inviscid and viscous flow at various time instants is shown in Figure 5. A perfectly spherical contour of the shock can be seen at time instant -24.3μs. After the shock reflects from the end wall, at time t=75.7 μs, a mushroom-shaped structure is observed behind the reflected shock for both cases. No noticeable effect of viscosity is seen before the shock reaches the focusing end wall. However, viscosity effects were observed after the shock reflection. The inviscid flow retains the spherical shape of the shock even after reflection. Whereas, for viscous flow, the shape of the reflected shock has become almost planar.

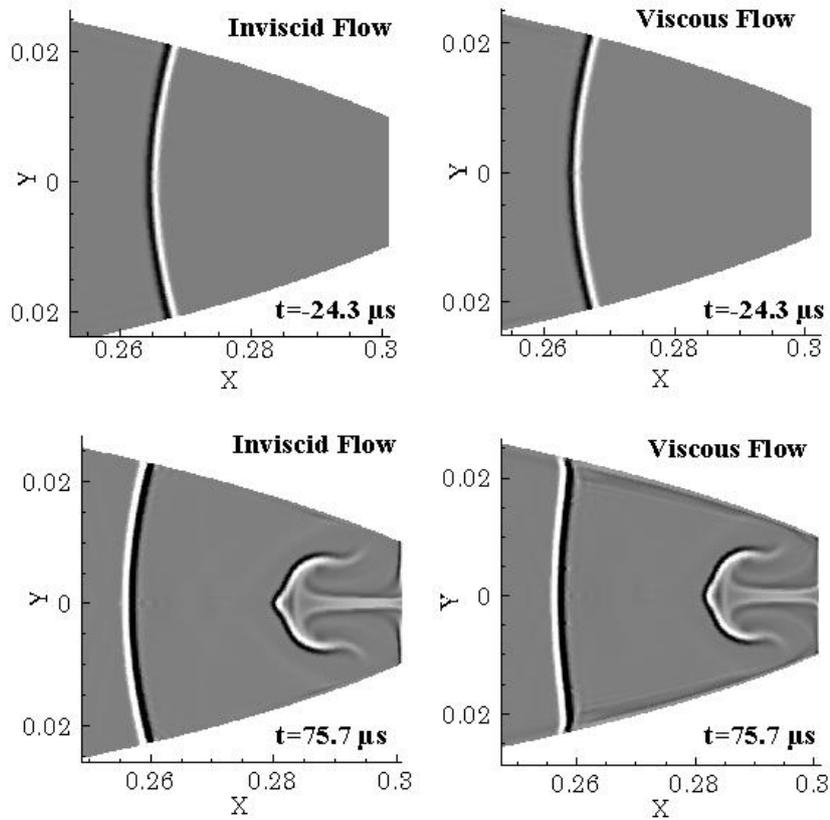

*Figure 5: Numerical shadowgraph images showing the comparison of inviscid and viscous flow for two time instants. One time instant is before the shock reaches focusing end wall and the other after the shock reflects from the end wall. The time t=0 corresponds to the time at which the shock reaches the end wall. The unit of measurement of x-axis and y-axis are in meters.*

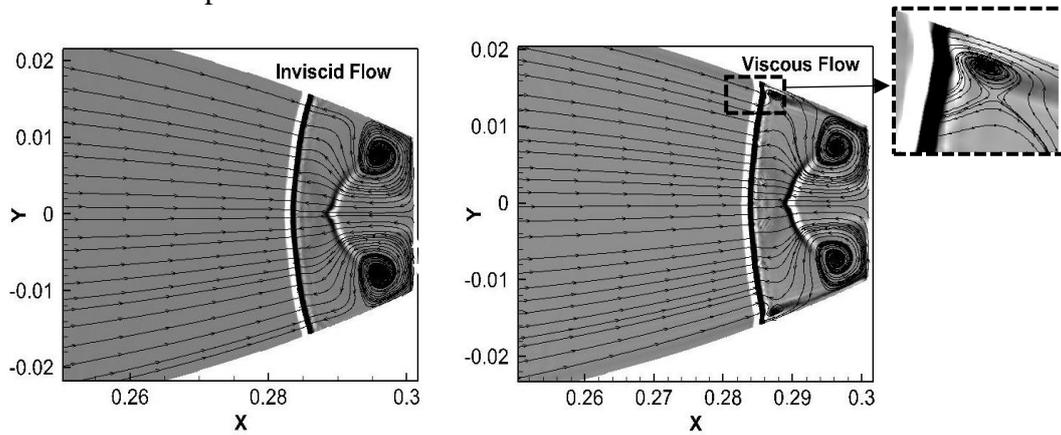

*Figure 6: Comparison of stream-traces of inviscid and viscous flow at an instant 31.7 μs*

For viscous flow case, the shock moves through an already developed boundary layer after reflection. This results in shock boundary layer interaction and associated flow separation locally, which in turn changes the overall shape of the shock from spherical to planar as it progresses upstream into the flow. This local separation and recirculation inside the boundary layer for the viscous flow case can be observed in the close-up image shown in Figure 6. Since the effect of viscosity is observed only towards the wall of the domain, further study is restricted to viscous results only.

In order to have an understanding of the mushroom structure formation, the numerical schlieren images at various instants after focusing are studied. Figure 7 shows the temporal evolution of the mushroom structure. The planar shock of initial strength $M_s$ =2.94 accelerates to a spherical shock of strength $M_s$ =4.5 when it reaches the focusing end wall. The high-energy spherical shock impinges on the focusing end wall and gets reflected, thereby resulting in a high-energy concentration region, high pressure, and temperature, as evident from the signals observed both experimentally and numerically. The reflected shock then travels upstream into the flow, as seen from the images in Figure 7. The high temperature and high-pressure gas slug then expand rapidly, as can be seen from time instants of 11.7 µs to 95.7 µs. This rapid expansion is similar to a jet blast at supersonic speeds [26] or an explosion occurring at open atmosphere. However, the only difference is here is that the flow is bounded on either side by the wall. The expanding hot gas slug is bounded by a curved wavefront, as seen at a time instant of 11.7 µs. At time instant of 31.7 µs after shock reflection at the end wall, a mushroom-shaped structure evolved.

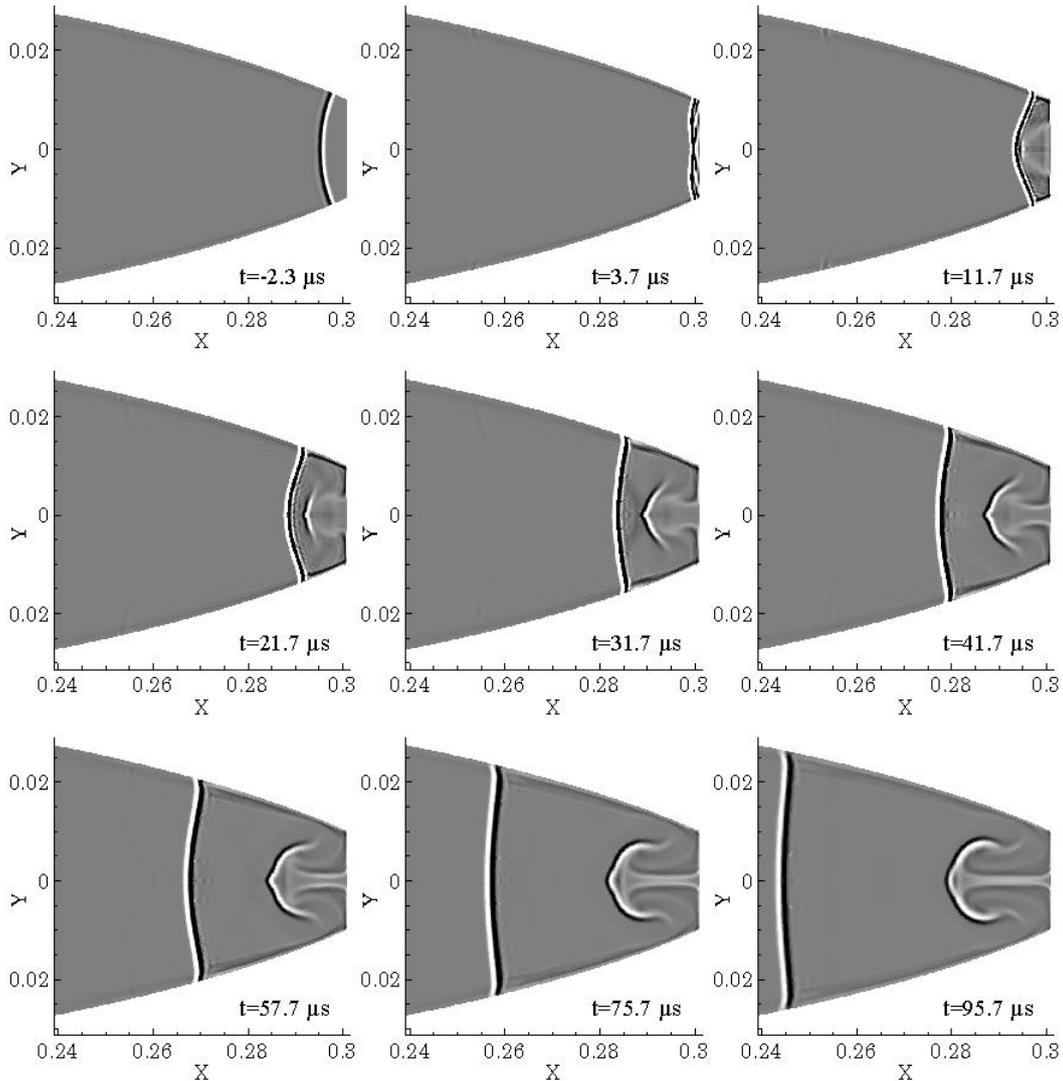

*Figure 7: Development of Mushroom structure*

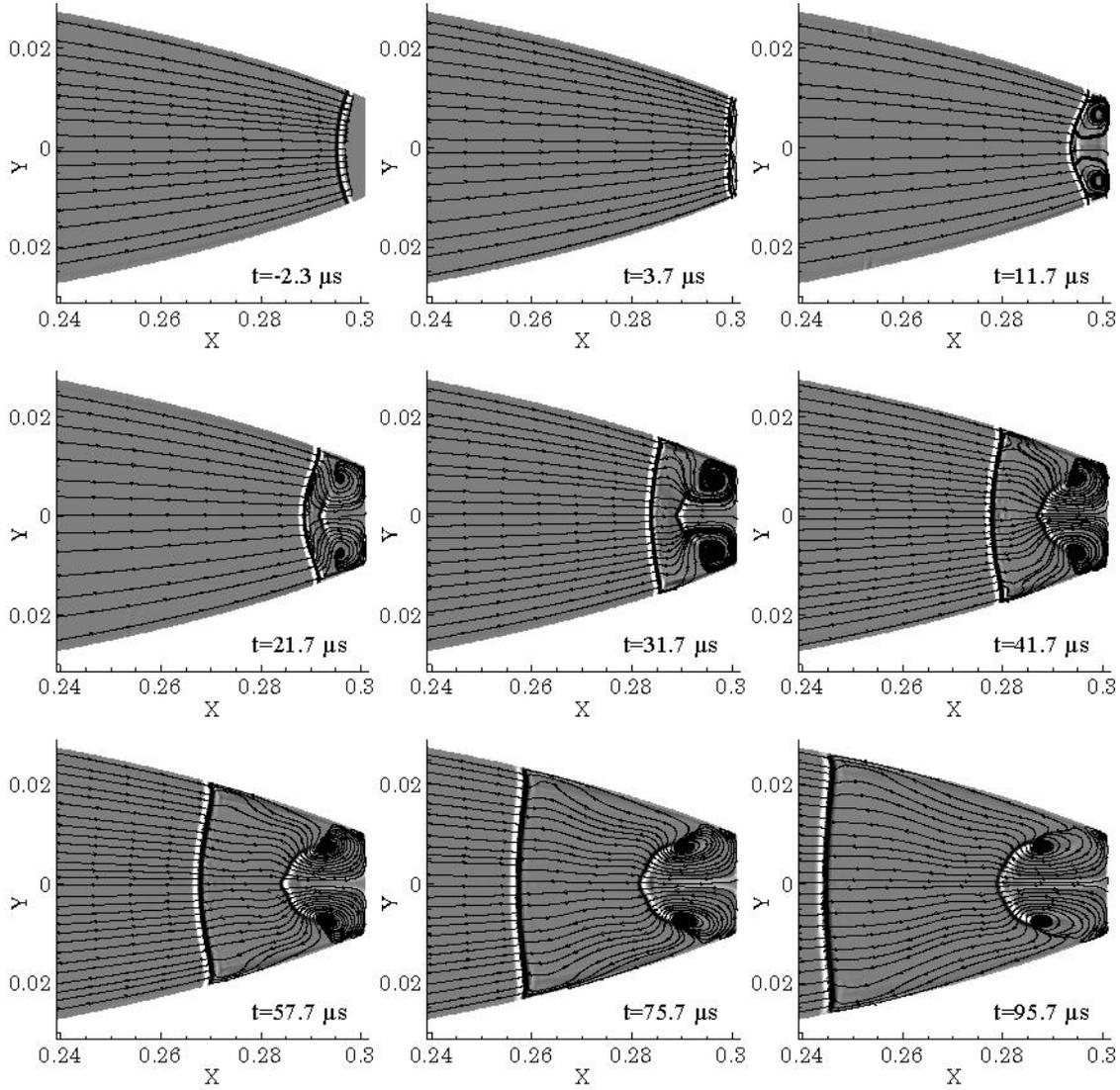

*Figure 8: The development of stream-traces and numerical shadowgraph over time*

A better understanding of the flow around the mushroom structure is made through stream-traces shown in Figure 8. Owing to the geometry, which is spherically diverging now, there exists a velocity gradient, in the expanding gas slug, across the axis, resulting in the generation of the vortical region as can be seen from the stream traces corresponding to time instant of 11.7 µs. The vortical region grows in size and strength, and its impingement on the end wall is what is observed as a second rise in the pressure signals obtained both experimentally and numerically. The second jump in pressure signal is observed between time 20 to 30 µs after the incident shock reflection at the end wall. This corresponds well with the stream-traces time instant images. At later time instants of 75.7 µs and 95.7 µs, we can see the vortical region getting elongated as the hot gas slug expands. The expanding hot gas slug finally takes the shape of a mushroom cloud, as can be seen from the figures. An animation showing the generation and growth of the mushroom structure is attached in the supplementary documents. (Open video).

Of more interest to us is to investigate the variation in thermodynamics properties at locations which is affected both by the passage of reflected shock and the expanding hot gas slug. The variation in temperature with time in the spherically contoured section is monitored at two different locations, x = 0.295 and 0.285, along with the end wall conditions, i.e., x=0.302. At x = 0.295, the arrival of the incident shock and reflected shock are clearly seen from the temperature signals depicted in Figure 9. The arrival of the mushroom head could be seen as a small and gradual rise in the temperature signal, the reason being the strength of the reflected shock at this location is stronger, and the temperature rise across it is of the same order as compared with the temperature inside the mushroom cloud region. For location x = 0.285, the arrival of incident shock, reflected shock, mushroom cloud, and Mach disc inside it are clearly

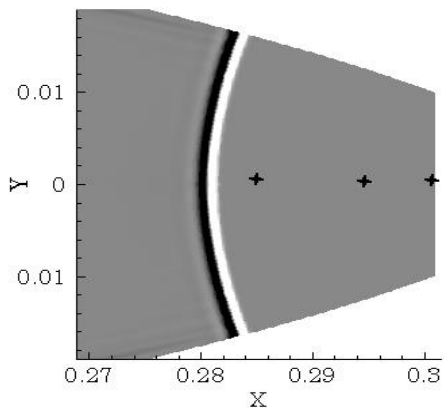

*Figure 10: Monitor points*

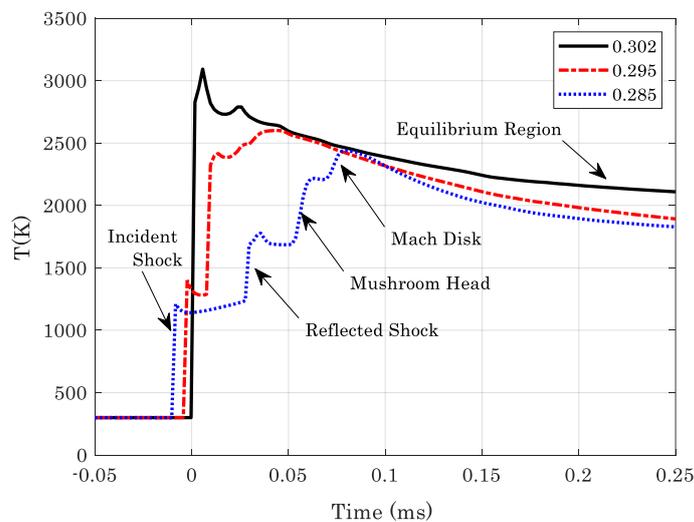

*Figure 9: Temperature monitored at three locations throughout the simulations.*

distinguishable as seen from the temperature signals. The reflected shock, which now sees a spherically diverging section, loses its strength which is clearly observed from the temperature rise behind it; nearly half the value when compared to the value at the location of x = 0.295. After a time instant of 100 µs, the temperature at both the locations was nearly the same, however, slightly lower than the temperature at the end wall. Of interest to note here is that the temperature in the expanding hot gas is high enough to cause vibrational excitation and possible dissociation of the gas molecules, thereby making the mushroom cloud region a chemically reacting zone. To understand if there was any dissociation of the gas

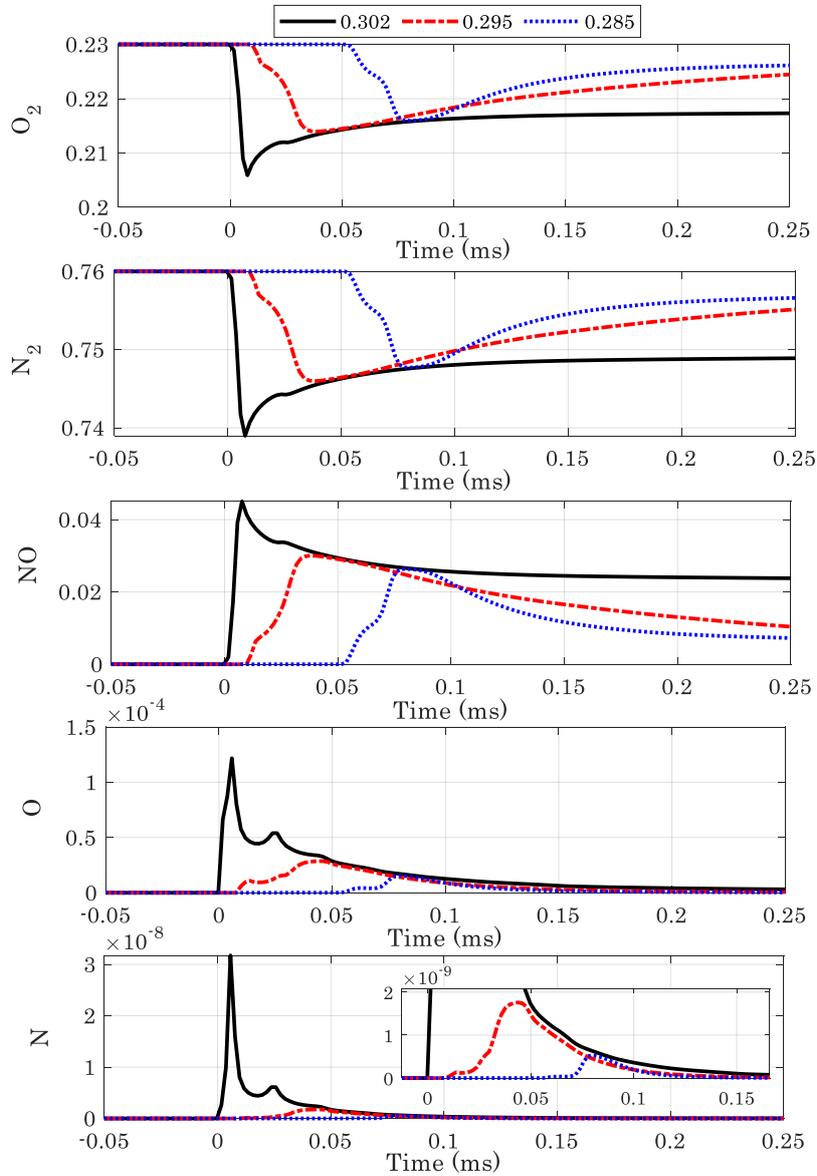

*Figure 11: Distribution of species mass fraction at three locations throughout the simulations.*

molecules, the species variation with time was examined and is plotted in Figure 11.

Dissociated mass fraction of species $N_2$ and $O_2$ was observed at both the chosen locations as well as at the end wall. $N_2$ and $O_2$ molecules dissociated and resulted in the formation of NO molecule, N, and O atoms. The species mass fraction of O and N atoms was relatively very low, of the order of 10 e$^{-4}$ and 10 e$^{-8}$, respectively. The dissociation was higher at the end wall location as expected, where the temperature is higher, and at the location of x = 0.295 and 0.285, it was relatively smaller. Nevertheless, these results indicate the expanding mushroom cloud is indeed chemically reacting, as can be seen from the plots. An animation showing the species mass fraction with corresponding shock position is mentioned in the supplementary material ([Open video](#)).

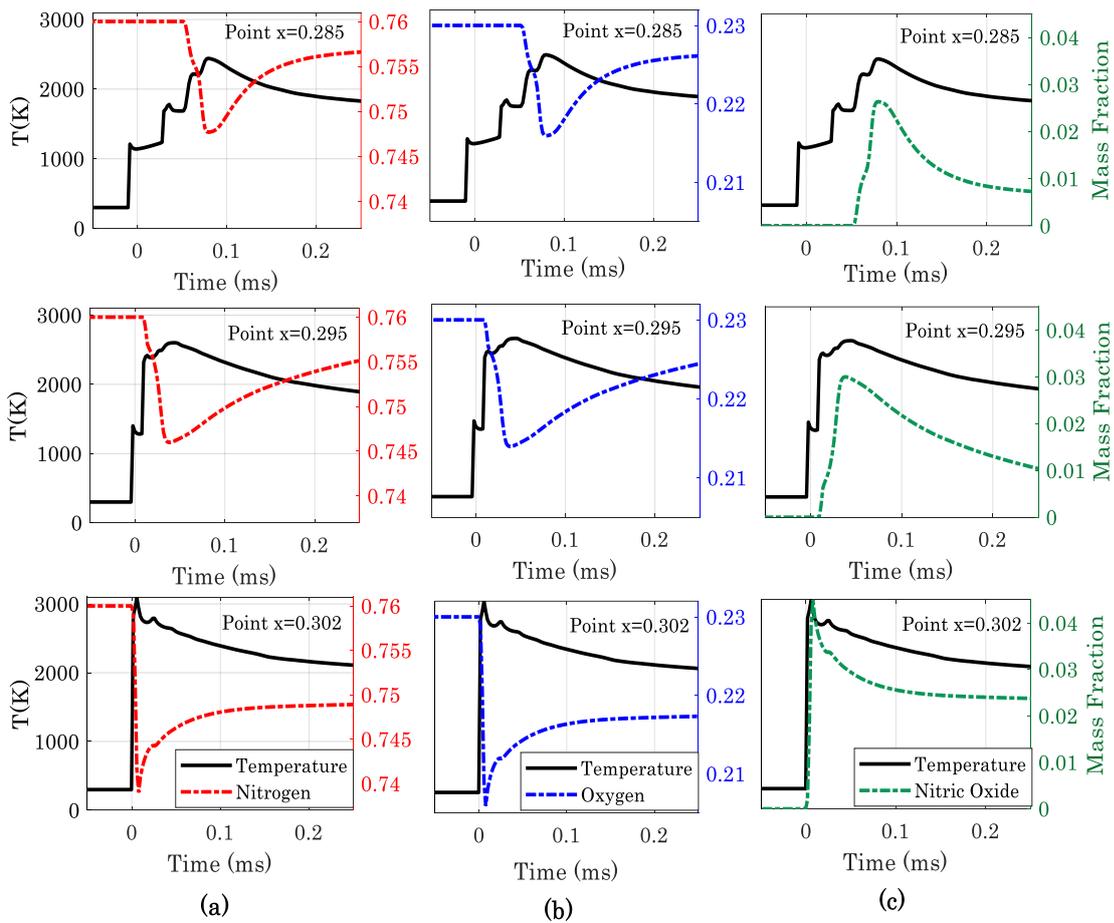

*Figure 12: The comparison of temperature distribution and mass fraction distribution of (a) Nitrogen, (b) Oxygen, and (c) Nitric Oxide at location x=0.285, 0.295, and 0.302.*

The variation in species mass fraction for $O_2$, $N_2$, and NO molecules corresponding to temperature change brought about by the passage of reflection shock and the mushroom cloud is plotted in Figure 12 for location x = 0.295, 0.285, and 0.302. for the x = 0.295 location, the dissociation of gas molecules starts after the reflected shock passes by, followed by the arrival of the chemically reacting mushroom cloud across which the species mass fraction changes further. However, for location x = 0.285, the passage of reflected shock did not result in dissociation of the gas molecules, as discussed earlier. The reflected shock loses its strength and thereby not causing any significant change in temperature as required for dissociation. It is with the arrival of the head of the mushroom cloud that dissociated mass fraction of $N_2$ and $O_2$ molecule and evolution of NO species were observed. At both these locations, the species mass fraction is continuously changing with time, owing to the continuous change in temperature, and finally attains an equilibrium value.

## 4   Conclusion

A shock of initial strength Ms=2.94 is focused to a confined region with the help of a smoothly converging section attached to a shock tube. The converging section converts the planar shock generated in the shock tube into a spherical shock with the least possible losses. Pressure is monitored inside the converging section, and these measurements show good agreement with numerical simulations. A further detailed study on the flow phenomenon in the focusing region is carried out with numerical simulations. A comparison of inviscid and viscous flow properties is also made through this evaluation. The effect of viscosity is found to be significant only near the wall due to the formation of boundary layer.

The shadowgraph images of the flow shows the presence of a mushroom shaped structure behind the reflected shock. After the shock reflects from the focusing end wall, the gas undergo expansion rapidly in a confined area, causing vortex formation, and as a result, a mushroom shaped structure is generated. The temperature distribution at three locations are monitored throughout the simulations and the effect of mushroom head, Mach disk inside the mushroom shaped structure etc. are clearly observed. The temperature inside the mushroom structure is found to be higher than that across the reflected shock. The distribution of the mass fraction of five significant species of air ($N_2$, $O_2$, $NO$, $N$, and $O$) are monitored. The species distribution across the mushroom structure shows that there is a reacting mixture of high-temperature gas inside this structure.

**Acknowledgment**

The authors would like to thank the Science and Engineering Research Board (SERB), India, for supporting this research work under the Early Career Research Award, ECRA/2018/000678.